\documentclass[conference]{IEEEtran}
\IEEEoverridecommandlockouts
\usepackage{cite}
\usepackage{amsmath,amssymb,amsfonts}
\usepackage{algorithmic}
\usepackage{algorithm}
\usepackage{graphicx}
\usepackage{textcomp}
\usepackage{xcolor}
\def\BibTeX{{\rm B\kern-.05em{\sc i\kern-.025em b}\kern-.08em
    T\kern-.1667em\lower.7ex\hbox{E}\kern-.125emX}}
\begin{document}

\title{Energy-Aware Multi-Agent Reinforcement Learning for Collaborative Execution in Mission-Oriented Drone Networks \\
}

\author{\IEEEauthorblockN{Ying Li\IEEEauthorrefmark{1}, Changling Li\IEEEauthorrefmark{1}, Jiyao Chen\IEEEauthorrefmark{2}, Christine Roinou\IEEEauthorrefmark{1}}
\IEEEauthorblockA{\IEEEauthorrefmark{1}Department of Computer Science, Colby College, Waterville, ME 04901, USA \\
\IEEEauthorrefmark{2}Department of Computer Science, Columbia University, New York, NY 10027, USA \\
Email:\{ying.li, changling.li\}@colby.edu}
}

\maketitle

\begin{abstract}
Mission-oriented drone networks have been widely used for structural inspection, disaster monitoring, border surveillance, etc. Due to the limited battery capacity of drones, mission execution strategy impacts network performance and mission completion. However, collaborative execution is a challenging problem for drones in such a dynamic environment as it also involves efficient trajectory design. We leverage multi-agent reinforcement learning (MARL) to manage the challenge in this study, letting each drone learn to collaboratively execute tasks and plan trajectories based on its current status and environment. Simulation results show that the proposed collaborative execution model can successfully complete the mission at least 80\% of the time, regardless of task locations and lengths, and can even achieve a 100\% success rate when the task density is not way too sparse. To the best of our knowledge, our work is one of the pioneer studies on leveraging MARL on collaborative execution for mission-oriented drone networks; the unique value of this work lies in drone battery level driving our model design.              
\end{abstract}


\section{Introduction}



The easy deployment and high mobility have made drones prevalent nowadays. Due to the potential of enhancing the network coverage and improving the efficiency of executing tasks in complex and dangerous environments,  drones have been used in various applications recently, such as using drones as mobile base stations in 5G networks~\cite{li2018uav} and mission-oriented drone networks for package delivery~\cite{choi2017optimization}, border surveillance~\cite{haddal2010homeland}, forest fire monitoring~\cite{kinaneva2019early}, etc.

Although drones can bring convenience in various scenarios, the limited battery capacity restricts the operation time, which is a challenging problem for researchers. With the current battery technology, drones can have approximately 20 to 40 minutes of operation time~\cite{hashemi2019new}. Researchers have explored two categories of approaches to overcome this challenge: battery or drone replacement and mission execution management. The first approach aims to design replacement mechanisms to let spare drones replace weary drones~\cite{9213932} or enable weary drones to recharge the batteries~\cite{9275880}. The mission execution management approach focuses on trajectory planning so that drones can efficiently finish their work before running out of power~\cite{liao2021trajectory, yao2019qos}.

However, mission execution management for multi-drone systems is still underdeveloped, with only a few works using centralized algorithms~\cite{li2021energy}. In real life, a mission is usually composed of multiple tasks, such as inspection of the wind turbines in a wind farm~\cite{barker2021semi}. The mission is composed of several tasks, and each is to examine a turbine. The energy spent on executing a task can vary depending on the task. Therefore, a single drone can finish some light tasks, while each of the other demanding ones may need multiple drones to execute cooperatively. Proper and robust execution management is necessary to efficiently finish the mission with the limited battery capacity and make drones more applicable in real life.

This paper proposes a collaborative execution model based on multi-agent reinforcement learning (MARL) and deep Q-network (DQN)~\cite{Mnih_jnature_2015} for mission-oriented drone networks to accomplish the assigned mission efficiently. We explore scenarios in which a mission comprises several independent tasks with various task lengths at different locations. A mission is considered to be completed when all of its tasks are finished. Each of these tasks needs one or more time steps to complete. A task with multiple time steps can have several discrete portions and can be co-finished by multiple drones, with each drone finishing a part of the task. Hence, a single drone may execute more than one task, and a task may be taken care of by more than one drone. Our study aims at collaborative task execution considering the battery level of drones and trajectory control, as less energy spent on travel from one task to another can let more energy be spent on task execution. To our best knowledge, our exploration is one of the pioneer studies focusing on collaborative mission execution while considering the energy consumption on travel using MARL for drone networks.

The rest of this paper is organized as follows. Section~\ref{related work} reviews the related work and compares their studies with ours to show our contribution to this field in multiple aspects. Section~\ref{system model} and ~\ref{proposed model} provide a detailed description of the mission-oriented drone network model we used in this study and the design of our collaborative execution model. Section~\ref{performance eval} describes our simulation environments, experiment design, simulation results. Section~\ref{conclusion} concludes and points out the future work of this study.

\section{Related Work}
\label{related work}



The methods proposed by existing studies can be categorized into two groups: methods without using machine learning and those leveraging machine learning. Most of these studies assume that tasks have binary lengths, which can be completed once drones pass through the locations where tasks are located. Therefore, these studies focus on optimizing trajectories to improve efficiency. 

Among the methods in the first category, a common way to establish the trajectory optimization models is \emph{waypoint segmentation}. Waypoint segmentation divides the trajectories based on different time intervals. Cobano~\textit{et al.} proposed RRT and RRT* algorithms to optimize the trajectories to address the Wireless Sensor Network (WSN) data collection problem defined by waypoint segmentation~\cite{cobano2013efficient}. The proposed RRT and RRT* algorithms focus on collision avoidance. Wang~\textit{et al.} considered the same problem in a larger scale network~\cite{wang2015efficient}. They divided the aerial data collection into five procedures, including waypoint searching, and proposed the FPPWR algorithm to increase the efficiency of path planning while assuring the relatively short trajectories. Qin~\textit{et al.} also explored the large-scale WSNs intending to minimize the completion time without compromising the information collection quality~\cite{qin2019completion}. They introduced a hovering point selection algorithm for appropriate waypoint selection and proposed a min-max cycle cover algorithm to allocate the  waypoints and compute the trajectories of each drone. 
These waypoint segmentation methods require the consideration of multiple segments for different drones during calculation at each time step, which can be computationally expensive. Xia~\textit{et al.} proposed to use time segmentation to establish the trajectory optimization model to simplify the calculation~\cite{xia2021multi}. Their proposed method reduces the computational complexity from $O(n^2)$  to $O(n)$ and is efficient for many drones. 

All these studies discussed above focus on collision avoidance while optimizing the trajectories. The task execution is either not considered or finished when the drones pass through the waypoints of tasks due to the binary task length. However, in reality, a task execution may take a more extended period to finish than what it needs for a drone to pass through a waypoint. Li~\textit{et al.} consider the scenario that the tasks have non-binary length~\cite{li2021energy}, similar to our proposed problem. They proposed the EATP model to accomplish all tasks energy efficiently. The advantage of these above non-machine-learning methods is that they guarantee an optimal or close-to-optimal solution. However, they usually require prior knowledge of the global information to calculate the paths and are less flexible to adapt to dynamic environments. Hence, they can be inefficient and time-consuming in reality applications. 

To combat the challenges of the dynamic environment, researchers have adopted machine learning in their solutions. Reinforcement learning (RL) has been prevalently used due to its nature of learning through interaction with the environment~\cite{sutton2018reinforcement}. Many researchers have employed RL to optimize drone networks' data sensing and transmission. The primary concern of Hu~\textit{et al.} is the probability of successful data transmission~\cite{hu2020reinforcement}. They employed Q-learning for trajectory control. To accelerate the Q-value convergence, they created an algorithm so that drones update their Q-functions based on the probability of successful data transmission. Fakhrul~\textit{et al.} concentrated on data freshness and energy efficiency~\cite{abedin2020data}. They proposed using the DQN with experience replay model to maximize the energy efficiency of the trajectories while assuring data freshness. A 4-dimensional numerical value represents the state space of their model. Other researchers have approached the state space differently. Bayerlein~\textit{et al.} proposed a DDQN model with convolutional layers for multi-drone path planning~\cite{bayerlein2021multi}. A combination of centralized global and local map representations of the environment is used as the state space. Their simulation shows that this approach enables the agents to divide the tasks and cooperate effectively. MADDPG~\cite{lowe2017multi} is another framework applied to multi-drone target assignment and path planning problems. 
Qie~\textit{et al.} proposed the STAPP method based on MADDPG to solve the task assignment and path planning problem~\cite{qie2019joint}. Their method can deal with the dynamic environments effectively as it only requires the information of the locations of drones, targets, and threat areas.

Similar to the methods discussed in the first category, those studies primarily focus on collision avoidance, trajectory planning, and task assignment. However, task lengths are out of consideration, so they didn't explore how to execute these non-binary tasks collaboratively.  
To bridge this gap, we propose a collaborative execution model driven by the battery level of drones, using MARL and deep Q-networks, which enables efficient mission execution. In summary, the contributions of this work are: 
\begin{itemize}
\item Unlike the previous work, we consider a practical scenario where a mission consists of multiple tasks with non-binary lengths. Each task requires multiple time steps to finish. To create a robust and applicable model, we also formulate the problem so that each task location and length of a mission are unpredictable.  
\item We leverage MARL and DQN to combat the challenge of the dynamic environment and solve the problems of collaborative execution. The drone battery level of all drones coupled with executed task portions drives the formulation of the reward function, enabling efficient cooperation. To our best knowledge, it is the first work that considers the constraints of battery capacity and task length in the model for drone networks.
\item Our simulation study is presented in detail to demonstrate that our model is robust and applicable. We also explore the effect of specific hyperparameters on the performance of our model and point out the challenges of applying MARL and our future research directions. 
\end{itemize}

\section{System Model}
\label{system model}
\begin{figure}[htbp]
\centerline{\includegraphics[width=\linewidth,height=0.5\linewidth]{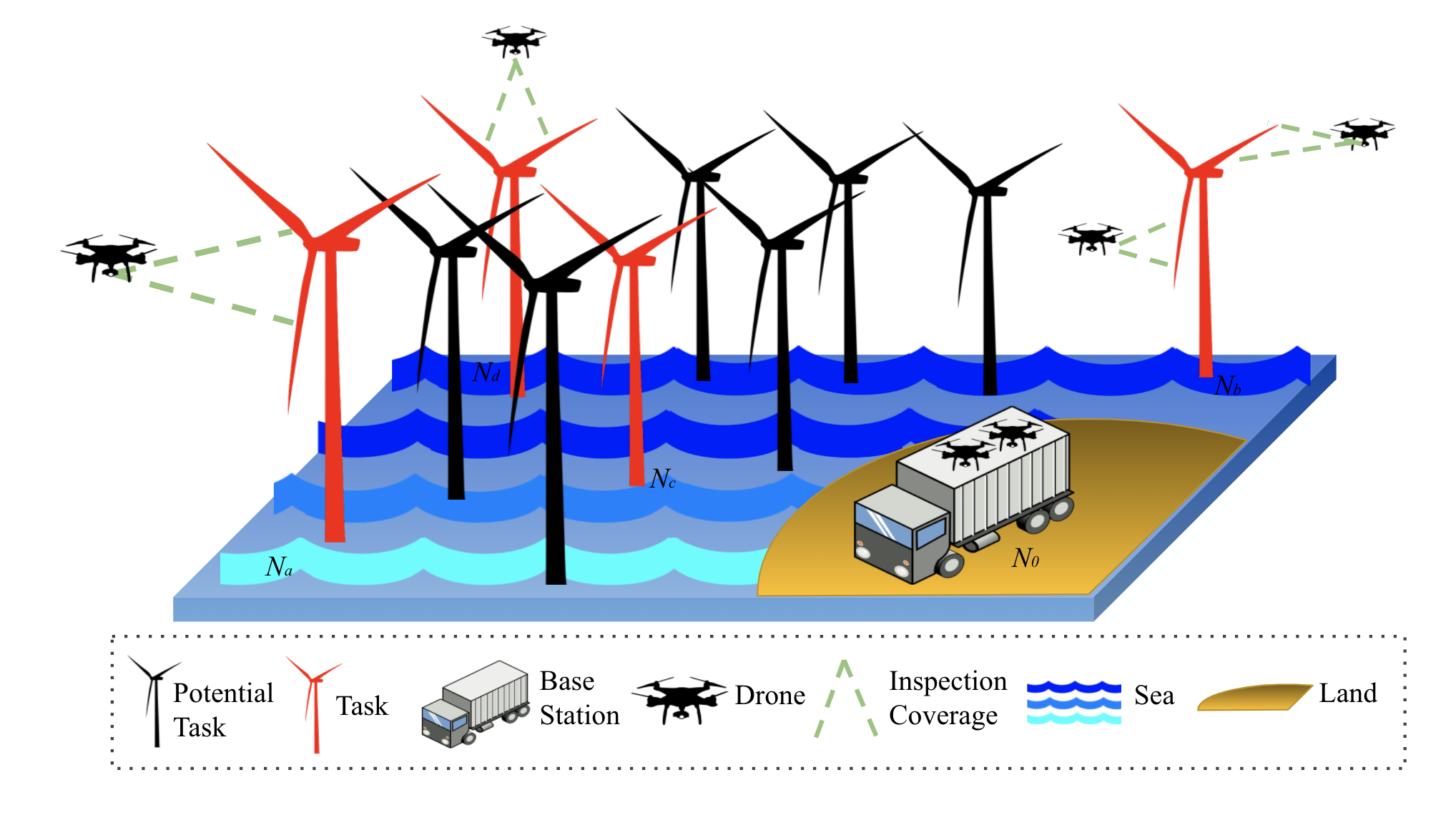}}
\caption{An example of the system model for mission-oriented drone networks and task execution in our study}
\label{fig1:system model}
\end{figure}

As illustrated in Fig.~\ref{fig1:system model}, we consider a set of trajectory points, $\mathcal{N}  = \{N_0, \dots, N_n\}$. The trajectory point $N_0$ is the location of the base station for drones, which is also the start point of all drone trajectories. Tasks are distributed at a subset of the remaining points. A task can locate at only one trajectory point, and a trajectory point can have at most one task. Therefore, the number of tasks assigned to the network is no more than $n$, as no task is located at the base station $N_0$. A task at the trajectory point $N_i, \forall i \in \{1, \dots, n\}$, needs $\mathcal{T}_i$ time steps to finish, $\mathcal{T}_i \geq 1$. This study considers that a mission comprises $K$ tasks distributed at $K$ different trajectory points, $1 \leq K \leq n$. Taking Fig.~\ref{fig1:system model} for instance, the mission comprises four tasks, $K = 4$, located at four out of ten trajectory points $n = 10$.

We assume in this study that a mission-oriented network of $K$ drones is assigned to execute a mission composed of $K$ tasks. All drones are fully charged and cooperatively complete the mission after launching from the base station $N_0$. Any fully charged drone in this study has the battery capacity $B$. A drone has to directly fleet to the trajectory point a task located from the drone's current location to execute it. A drone can decide how many time steps of a task it wants to execute based on its current status and current environment. Therefore, a drone can execute a portion of a task, a few time steps of $\mathcal{T}_i$, and multiple drones can co-finish a task. We assume that portions do not physically overlap in this study. For example, each portion is a part of a turbine to be examined. If all tasks are finished, drones return to the base station directly from their current location. Otherwise, drones move to the unfinished tasks and execute them.   

The energy consumption of a drone to execute a task per time step is composed of two parts: a part used to support the drone to hover at that trajectory point and the other consumed by the facilities (e.g., camera, sensors, etc.) carried by the drone to execute the task. In addition, flying forward consumes drone energy. We follow the method proposed by Stolaroff~\textit{et al.}~\cite{Stolaroff_nature_2018} to calculate the energy consumption rate for a drone hovering per time step, $P_{hover}$, defined by Equations~\ref{eq:phover},

\begin{equation}
P_{hover} = \frac{(m_{drone} g + F_{drag})^\frac{3}{2}}{\eta\sqrt{\frac{1}{2}\pi cD^2\rho}},
\label{eq:phover}
\end{equation}

\noindent where $m_{drone}$ is the mass of a drone, including its battery and facilities equipped on it,  $g$ is the gravitational constant, $F_{drag}$ is the drag force, $\eta$ is the overall power efficiency, $c$ is the number of rotors a drone has, the diameter of each rotor is $D$, and $\rho$ is the density of air. By using the average ground speed of a drone, $v$, and the pitch angle for steady flight, $\alpha$, the estimated minimum energy consumption rate for a drone flying forward per time step, $P_{forward}$ is defined by Equation~\ref{eq:forward},

\begin{equation}
P_{forward} = \frac{(m_{drone} g + F_{drag})(v \sin\alpha + v_s)}{\eta},
\label{eq:forward}
\end{equation} 

\noindent where $v_s$ is the included velocity to achieve the thrust to forward at a desired speed and can be found by the solution of Equation~\ref{eq:vs},

\begin{equation}
v_s = \frac{2 (m_{drone} g + F_{drag})}{\pi n D^2 \rho \sqrt{(v \cos\alpha)^2+(v \sin\alpha + v_s)^2}}.
\label{eq:vs}
\end{equation}

The energy consumption rate for the facilities on a drone used for task execution per time step is $P_{facilities}$ depends on the facilities carried by a drone. We assume that all drones have the same configuration, carry the same facilities to execute tasks, and travel at a uniform speed for calculation simplicity in this study. So, $P_{facilities}$, $P_{hover}$, and $P_{forward}$ remain the same for all drones.

The ultimate goal of a mission-oriented drone network is to achieve the mission assigned to the network. The mission is considered to be completed when two requirements are fulfilled: 1) all tasks of the mission are finished, and meanwhile 2) all drones can return to the base station before running out of battery. These two constraints are described by the following formulae~\ref{eq:constraint1} and~\ref{eq:constraint2}.

\begin{equation}
\sum_{i=1}^{K} \tau_i = 0, 
\label{eq:constraint1}
\end{equation} 
\begin{equation}
B_k \geq B^k_{return},\forall k \in \mathcal{K},
\label{eq:constraint2}
\end{equation}

\noindent where $\tau_i$ indicates the remaining time steps of the task at trajectory point $N_i$, $B_k$ is the current battery level of the drone $k$, $B^k_{return}$ is the amount of energy needed for the drone $k$ to return to the base station from its current location, and $\mathcal{K}  = \{1, \dots, K\}$. The valid value range for $\tau_i$ is $0 \leq \tau_i \leq \mathcal{T}_i$, where $\mathcal{T}_i$ is the original length of the task at trajectory point $N_i$. The valid value range for $B_k$ is $B_k \leq B$. $B_k$ can be negative values in our system model, indicating how much energy a drone spent beyond the battery capacity $B$. Any $B_k < B^k_{return}$ indicates that a mission is not successfully completed even if all tasks are finished, as the drone $k$ doesn't have enough energy to return to the base station. Our proposed collaborative execution model trains the drones to learn to finish all tasks efficiently without draining the battery.

$B^k_{return}$ is calculated using Equation~\ref{eq:b_return},
 
\begin{equation}
B^k_{return} = \frac{d(N_i, N_0) \times P_{forward}}{v},
\label{eq:b_return}
\end{equation}

\noindent where $d(N_i, N_0)$ is the Euclidean distance between the current location $N_i$ of the drone $k$ and the base station $N_0$. Please note that the returning drones don't execute any tasks on the way back to the base station.

We also assume that drones in our study are intelligent like the new Skydio 2 that can automatically avoid obstacles and collisions.


%

\section{Proposed Collaborative Execution Model Based on MARL and DQN}
\label{proposed model}


We propose a collaborative execution model based on MARL and DQN to address the limited battery capacity when executing the mission assigned to a drone network. Our model lets each drone have a DQN to find the best trajectory and task execution strategy to successfully co-finish the mission with other drones. All drones in this model share a reward function to collaborate with others fully.

Every DQN in our study consists of a policy network and a target network. These two networks have two fully-connected hidden layers using the \emph{rectified linear unit} activation function. They also have a densely-connected output layer. The loss function used in this DQN model is the \emph{mean squared error}, and we adopt the \emph{Adam} algorithm to optimize speed and performance for training the model. We also combine our DQN with \emph{experience replay} to break the dependency among the observations in the training process.

\subsection{Action and State Space}
\begin{figure}[htbp]
\centerline{\includegraphics[scale=0.35]{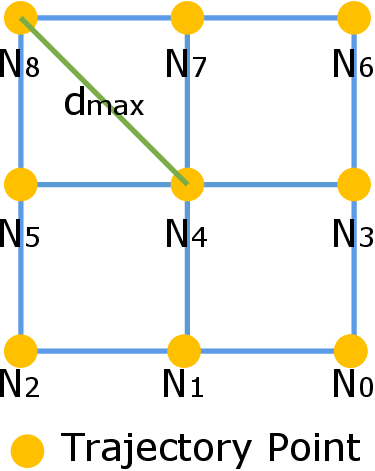}}
\caption{A grid pattern for nine trajectory points}
\label{fig:gridPattern}
\end{figure}

Inspired by~\cite{Hu_jiot_2018}, we divide the space into a finite set of discrete trajectory points $\mathcal{N}$ in a grid pattern. Each time step $t, t \in \{0, \dots, T\}$, a drone can travel from a trajectory point $N_{start}$ to another trajectory point $N_{end}$ on the same square, $N_{start} \in \mathcal{N}, N_{end} \in \mathcal{N}$. The maximum distance $d_{max}$ a drone can travel in a time step is the diagonal between two trajectory points located on the same square. When the travel distance $d(N_{start}, N_{end})< d_{max}$, the drone will hover at the endpoint for the rest of the current time step. Fig.~\ref{fig:gridPattern} shows a sample grid pattern with nine trajectory points. At any time step, a drone can travel from $N_4$ to any one of its eight adjacent trajectory points. It takes exactly one time step for a drone to travel from $N_4$ to any one of \{$N_0$,  $N_2$,  $N_6$,  $N_8$\}. The drone will hover at the endpoint for the rest of the current time step if it travels from $N_4$ to any one of \{$N_1$,  $N_3$,  $N_5$,  $N_7$\}.

The set of actions $\mathcal{A}$ available to the drones consists of ten actions, which can be grouped into three types. The first type is to let a drone move from one trajectory point to another, and the travel distance per time step for any action of this type fulfills $d(N_i, N_j) \leq d_{max}$. Eight actions belong to this type, allowing a drone to move to the adjacent trajectory points in eight different directions: up, down, right, left, upper left, upper right, bottom left, and bottom right. The first type results in a new trajectory point for a drone before the beginning of the next time step and consumes the drone's power for travel. The second type contains one action, letting a drone hover at the current trajectory point, $N_i$, if $N_i \in \{N_1, \dots, N_n\}$. If the current trajectory point is $N_0$, the drone stays stationary at the base station without consuming any energy. The second type doesn't update a drone's location but consumes the drone's power for hovering when the drone is not at the base station. The third type has one action, allowing a drone to execute the task at the trajectory point the drone currently locates while hovering there. This type doesn't change a drone's location but consumes the drone's energy for task execution and hovering. Each drone $k$ chooses an action $a$ from $\mathcal{A}$ at the beginning of each time step $t$ and then updates its location and battery level $B_k$ accordingly.    

The state space for the proposed collaborative execution model is a quintuple. Each time step $t$, the state space is denoted by Equation~\ref{eq:statespace},

\begin{equation}
S_t=\langle L_{task}^{\mathcal{K}}, L_{drone,t}^{\mathcal{K}}, A_{t}^{\mathcal{K}}, \tau_{t}^{\mathcal{K}}, B_{t}^{\mathcal{K}} \rangle,
\label{eq:statespace}
\end{equation} 

\noindent where $L_{task}^{\mathcal{K}}$ contains the locations of all $K$ tasks, $L_{drone,t}^{\mathcal{K}}$ contains the current locations of all $K$ drones, $A_{t}^{\mathcal{K}}$ contains the current action of each of the $K$ drones, $\tau_{t}^{\mathcal{K}}$ contains the remaining time steps of each of the $K$ tasks, $B_{t}^{\mathcal{K}}$ contains the current battery levels of each of the $K$ drones, and they are defined by the following Equations~\ref{eq:L_task_k constraint} to~\ref{eq:B_t_k constraint}.

\begin{equation}
L_{task}^{\mathcal{K}}  = \{l_{task}^k | l_{task}^k \in \mathcal{N} \backslash \{N_0\}, \forall k \in \mathcal{K}\},
\label{eq:L_task_k constraint}
\end{equation} 

\begin{equation}
L_{drone,t}^{\mathcal{K}}  = \{l_{drone,t}^{k} | l_{drone,t}^{k} \in \mathcal{N}, \forall k \in \mathcal{K}\},
\label{eq:L_drone,t_k constraint}
\end{equation} 
 
\begin{equation}
A_{t}^{\mathcal{K}} = \{a_{t}^{k} | a_{t}^{k} \in \mathcal{A}, \forall k \in \mathcal{K}\},
\label{eq:A_t_k constraint}
\end{equation} 

\begin{equation}
\tau_{t}^{\mathcal{K}} = \{\tau_{t}^{k} | 0 \leq \tau_{t}^{k} \leq \mathcal{T}_k, \forall k \in \mathcal{K}\},
\label{eq:tau_t_k constraint}
\end{equation} 

\begin{equation}
B_{t}^{\mathcal{K}} = \{B_{t}^{k} | B_{t}^{k} \leq B, \forall k \in \mathcal{K}\}.
\label{eq:B_t_k constraint}
\end{equation}


\subsection{Reward and Control Policy}
At the beginning of each time step, every drone chooses an action that transmits the environment state from $S_t$ to $S_{t+1}$ and generates an immediate reward $R_{t+1}$. The reward is then used to form the control policy that drives the drones to accomplish the mission efficiently by optimizing the accumulated reward. We expect drones to execute tasks collaboratively while being aware of their battery levels. Hence, the reward is formulated based on the execution progress of all tasks and the remaining battery power of all drones. The reward function is shared among the drones, which encourages drones to collaborate as a team. The task execution progress per time step, $E(t)$, is defined by Equation~\ref{eq:taskexecution},

\begin{equation}
E(t) = \sum_{k=1}^{K}(\tau^k_{t-1}-\tau^k_{t}),
\label{eq:taskexecution}
\end{equation} 

\noindent and the reward function is defined by Equation~\ref{eq: reward},

\begin{equation}
\resizebox{0.48\textwidth}{!}{$
    R_{t+1}= 
\begin{cases}
    E(t+1) + \gamma \mu_{t+1},              & \text{if formulae~\ref{eq:constraint1} and~\ref{eq:constraint2} are true,}\\
    E(t+1) - \beta \omega_{t+1},              & \text{if formula~\ref{eq:constraint1} is true and formula~\ref{eq:constraint2} is false}, \\
    E(t+1),                                             & \text{if formula~\ref{eq:constraint1} is false,}
\end{cases}
$}
\label{eq: reward}
\end{equation} 

\noindent where $\mu_t$ measures the rate of the total amount of remaining energy of all drones at time step $t$, defined by Equation~\ref{eq:energyleft}, 

\begin{equation}
\mu_t = \frac{\sum_{k=1}^{K}B^k_t}{KB}, 
\label{eq:energyleft}
\end{equation} 

\noindent $\omega_t$ tells the number of drones that run out of power at the time step $t$, shown by Equation~\ref{eq:numdeaddrones},
 
\begin{equation}
\omega_t = |\{B^k_t \mid \forall k \in \mathcal{K}, B^k_t < B_{return}\}|, 
\label{eq:numdeaddrones}
\end{equation} 

\noindent and $\gamma$ and $\beta$ are coefficients used to adjust the impact of $\mu_t$ and $\omega_t$ to achieve better energy efficiency. This reward function encourages drones to execute more tasks or more task portions per time step while reserving energy and discourages them from using up their battery for unnecessary travel or inefficient cooperation.

\begin{algorithm}[htbp]
\caption{Energy-Aware Multi-Agent DQN for Collaborative Task Execution}
\label{alg:strategyoutline}
\begin{algorithmic}[1] 
\FOR{Each drone $k = 1, \dots, K$}
  \STATE Initialize a replay memory, a policy network, and a target network 
\ENDFOR
\FOR{Each episode}
    \STATE Initialize $S_0$
    \FOR{$t = 0, \dots ,T$} 
        \FOR{Each drone $k = 1, \dots, K$}
        	  \STATE Generate a random probability 
        	   \IF { The probability $\leq \Delta$}
	   	\STATE Choose an action from $\mathcal{A}$ via exploration
	   \ELSE
	   	\STATE  Choose an action from $\mathcal{A}$ via exploitation
	   \ENDIF
            \STATE Execute the selected action 
            \STATE Observe $R_{t+1}$ and $S_{t+1}$
            \STATE Store $\{S_t, A^k_t, R_{t+1}, S_{t+1}\}$ in the replay memory of drone $k$
            \IF {The number of stored samples $> batch\_size \times \psi$}
            	\STATE Update the weights in the target network to the weights in the policy network every $\chi$ time steps.
            	\STATE Sample a random batch from the replay memory of drone $k$
		\STATE Pass the batch of states to the policy network
		\STATE Pass the batch of states to the target network
		\STATE Calculate the loss between the output Q-values of the policy network and the Q-values of the target network
		\STATE Updates weights in the policy network to minimize the loss
            \ENDIF
        \ENDFOR
     \ENDFOR
\ENDFOR
\end{algorithmic}
\end{algorithm}

The action and state space coupled with the reward function are used in the proposed collaborative execution model, described as Algorithm~\ref{alg:strategyoutline}. This proposed model lets each drone have its DQN and individually learn based on the share reward function and state space. We set a threshold, $batch\_size \times \psi$, in this model to allow it to start learning from the past experience when there are enough samples stored in the replay memory with the consideration of avoiding unnecessary instability. In this model, another threshold, \emph{exploration rate} $\Delta$, is to control the probability the model chooses an action based on its learning. If the probability is not larger than $\Delta$, the model randomly picks an action from $\mathcal{A}$ and investigates what occurs in the environment. Otherwise, the model selects the action with the highest Q-value for its present state. We will explore these two thresholds more in the next section.

\section{Performance Evaluation}
\label{performance eval}

We adopt the base case parameters published in~\cite{Stolaroff_nature_2018} to calculate the energy consumption rates in our simulation, as shown in Table~\ref{tab:parameter}.

\begin{table}[htbp]
\caption{Base Case Parameters to Calculate Energy Consumption Rates}
\label{tab:parameter}
\begin{tabular}{cccccc}
\hline
 \textit{$D$[m]} & \textit{$m_{drone}$[kg]} & \textit{$v$[m/s]}  & \textit{$\alpha$[rad]} & \textit{$\eta$} & \textit{$\rho$[$kg/m^2$]}    \\  [0.5ex]
\hline
\hline
 0.254  & 2.07  & 10 & 0.0139  & 0.7 &  1.2193\\ [0.5ex]
\hline
\end{tabular}
\end{table}

After getting the values, we proportionally scale down the energy consumption rate values calculated based on the base case parameters to $P_{facilities} = 3$, $P_{hover} = 4$, and $P_{forward} = 2.5$ for calculation simplification in our experiments. We set the battery capacity $B = 1800$ so that the drones would have enough energy to learn to execute tasks cooperatively. The DQNs of drones are built by using Tensorflow-Keras.

This study uses the \emph{success rate} to measure the performance of the proposed energy-aware collaborative execution MARL model. The success rate is calculated as the ratio between the number of accomplished missions and the total number of missions during a specific period. The \emph{average accumulated reward of successful missions} is adopted to measure the execution efficiency of completed missions. The accumulated reward is the sum of rewards of all steps of an episode. The average accumulated reward of successful missions is the mean of the accumulated rewards of episodes in which the mission has been successfully completed. A higher average accumulated reward of successful missions indicates that drones finish tasks more quickly, and their battery level is also high when the missions are accomplished. It also indicates that drones collaborate well and spend less power on unnecessary travel.

\begin{figure}[htbp]
\centerline{\includegraphics[scale=0.25]{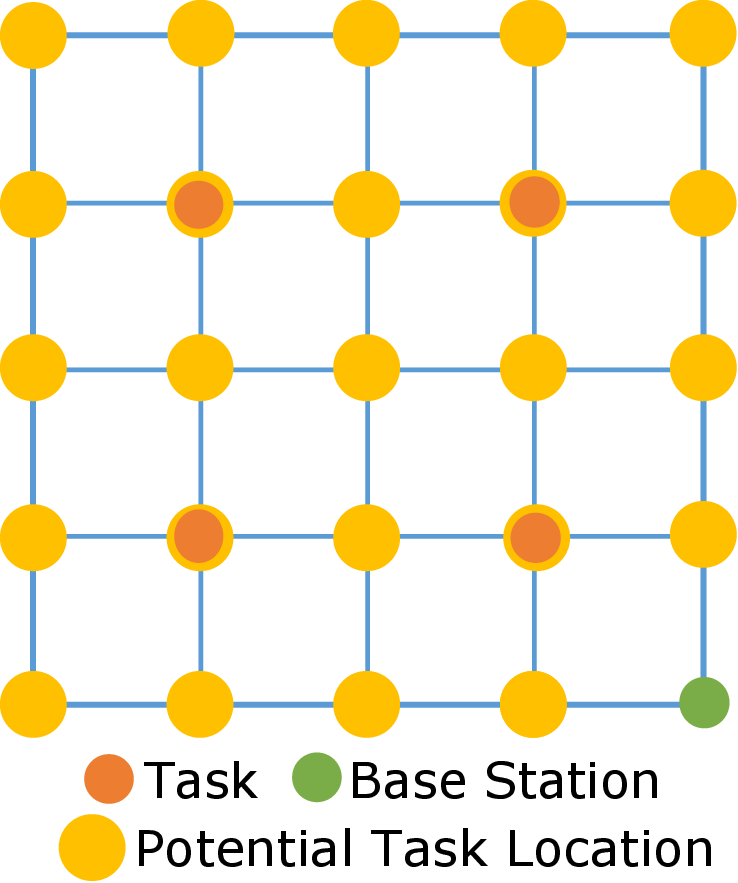}}
\caption{A $5 \times 5$ grid pattern consisting of 25 trajectory points. It shows a scenario of four tasks coupled with the base station on this grid.}
\label{fig:4 tasks}
\end{figure}

We conducted three sets of experiments in this study. The first set is to study the impact from the two thresholds, $\psi$ and $\Delta$, of the proposed model. Inspecting how the task length and location impact the model's performance is the goal of the second set of experiments. The last set investigates whether the task density affects the performance. We leverage a $5 \times 5$ grid pattern for all three experiment sets, consisting of 25 trajectory points, as shown in Fig.~\ref{fig:4 tasks}. We set the length of each episode to 600 time steps for all three sets of experiments.

\begin{figure*}[htp]
\begin{minipage}[t]{.5\linewidth}
\centerline{\includegraphics[width=\linewidth]{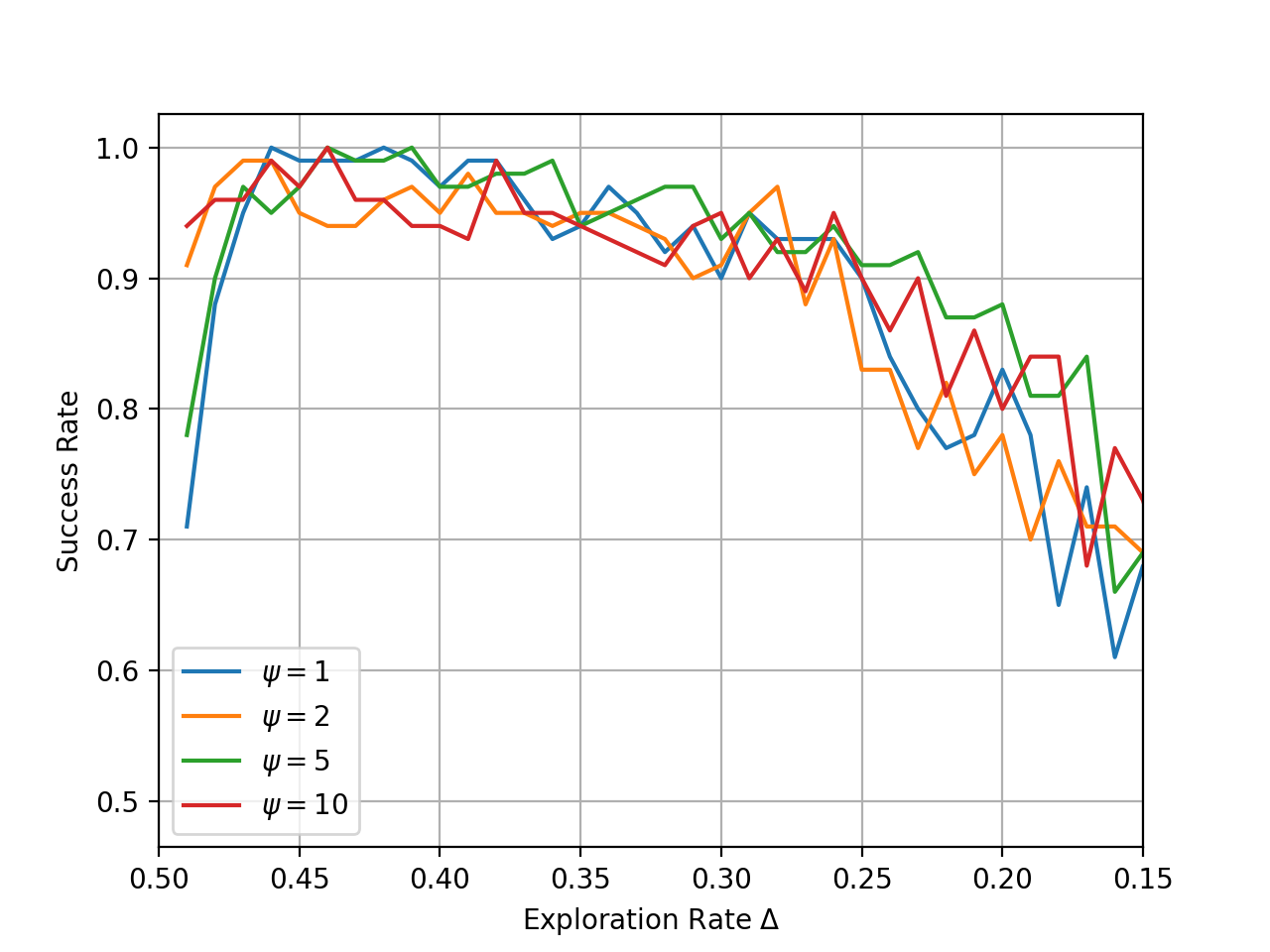}}
\begin{center}
\small \emph{(a) The success rates under different $\psi$ values and exploration rates}
\end{center}
\end{minipage}
\hfill
\begin{minipage}[t]{.5\linewidth}
\centerline{\includegraphics[width=\linewidth]{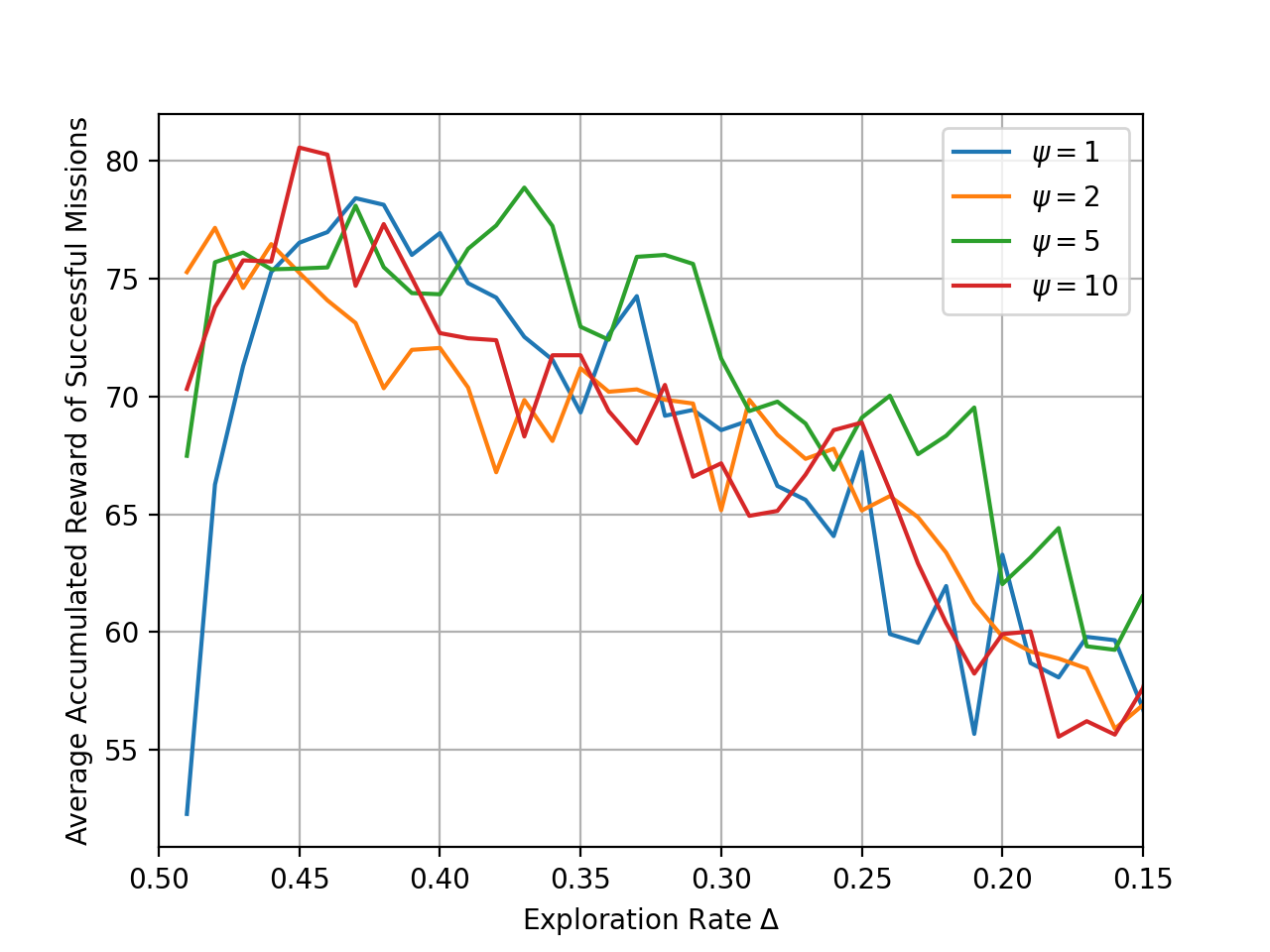}}
\begin{center}
\small \emph{(b)  The average accumulated rewards of successful missions under different $\psi$ values and exploration rates}
\end{center}
\end{minipage}
\caption{Results for the threshold experiments}
\label{fig:threshold exp}
\end{figure*}

\begin{figure*}[htp]
\begin{minipage}[t]{.5\linewidth}
\centerline{\includegraphics[width=\linewidth]{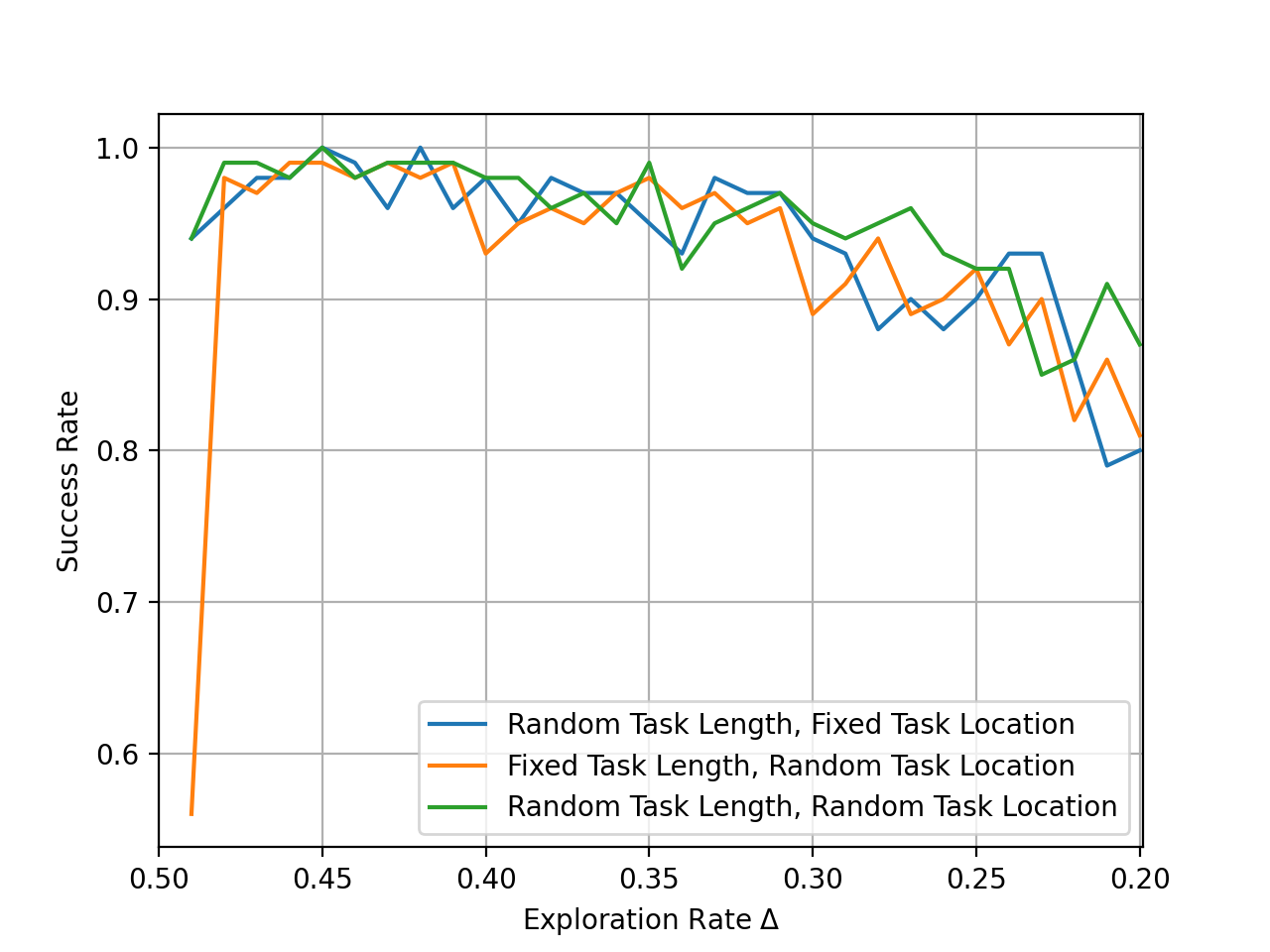}}
\begin{center}
\small \emph{(a) The success rates when the task length and task location are random}
\end{center}
\end{minipage}
\hfill
\begin{minipage}[t]{.5\linewidth}
\centerline{\includegraphics[width=\linewidth]{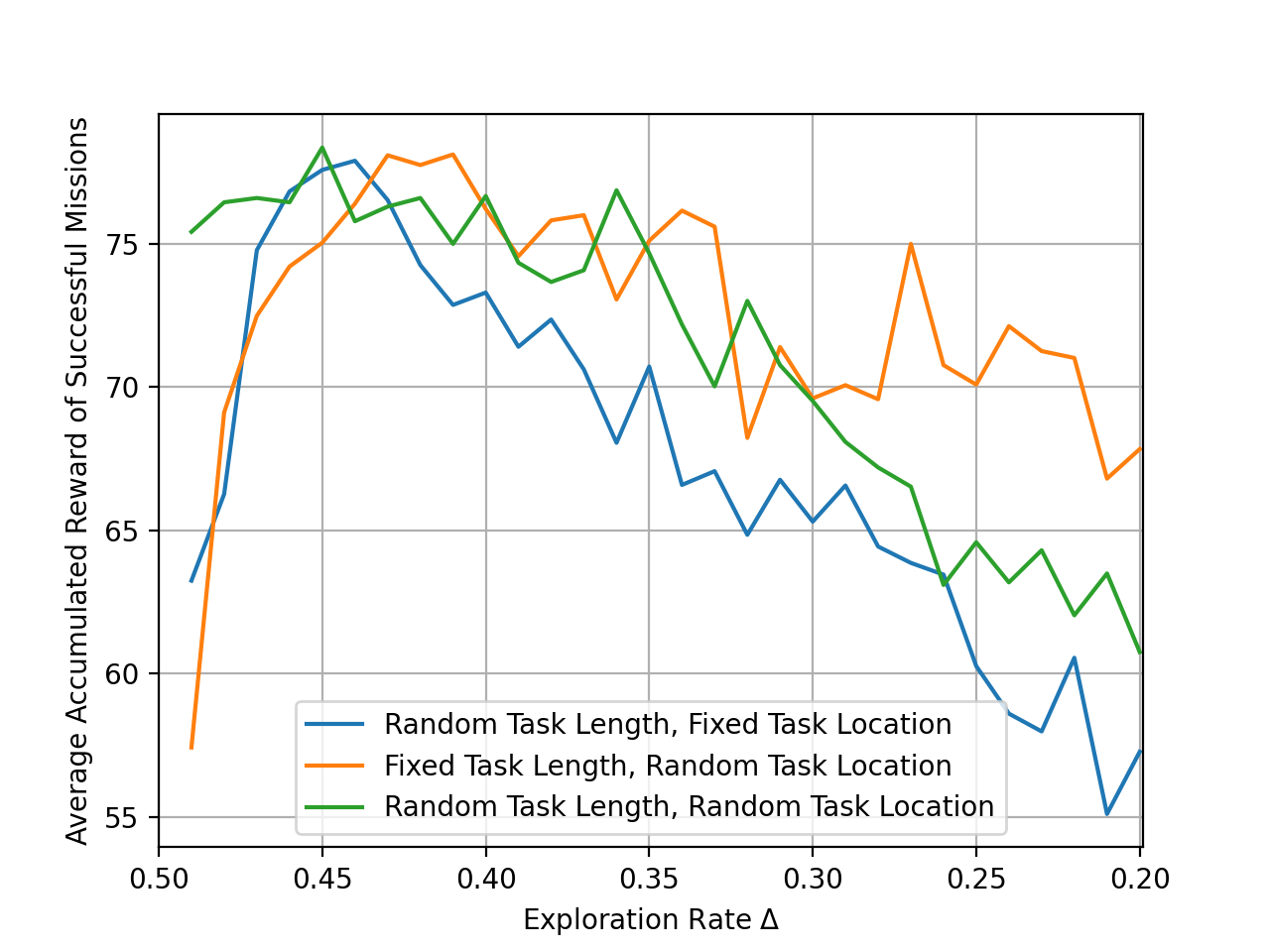}}
\begin{center}
\small \emph{(b) The average accumulated rewards of successful missions when the task length and task location are random}
\end{center}
\end{minipage}
\caption{Results for the task length and location experiments}
\label{fig:task len and loc exp}
\end{figure*}

\begin{figure*}[htp]
\begin{minipage}[t]{.5\linewidth}
\centerline{\includegraphics[width=\linewidth]{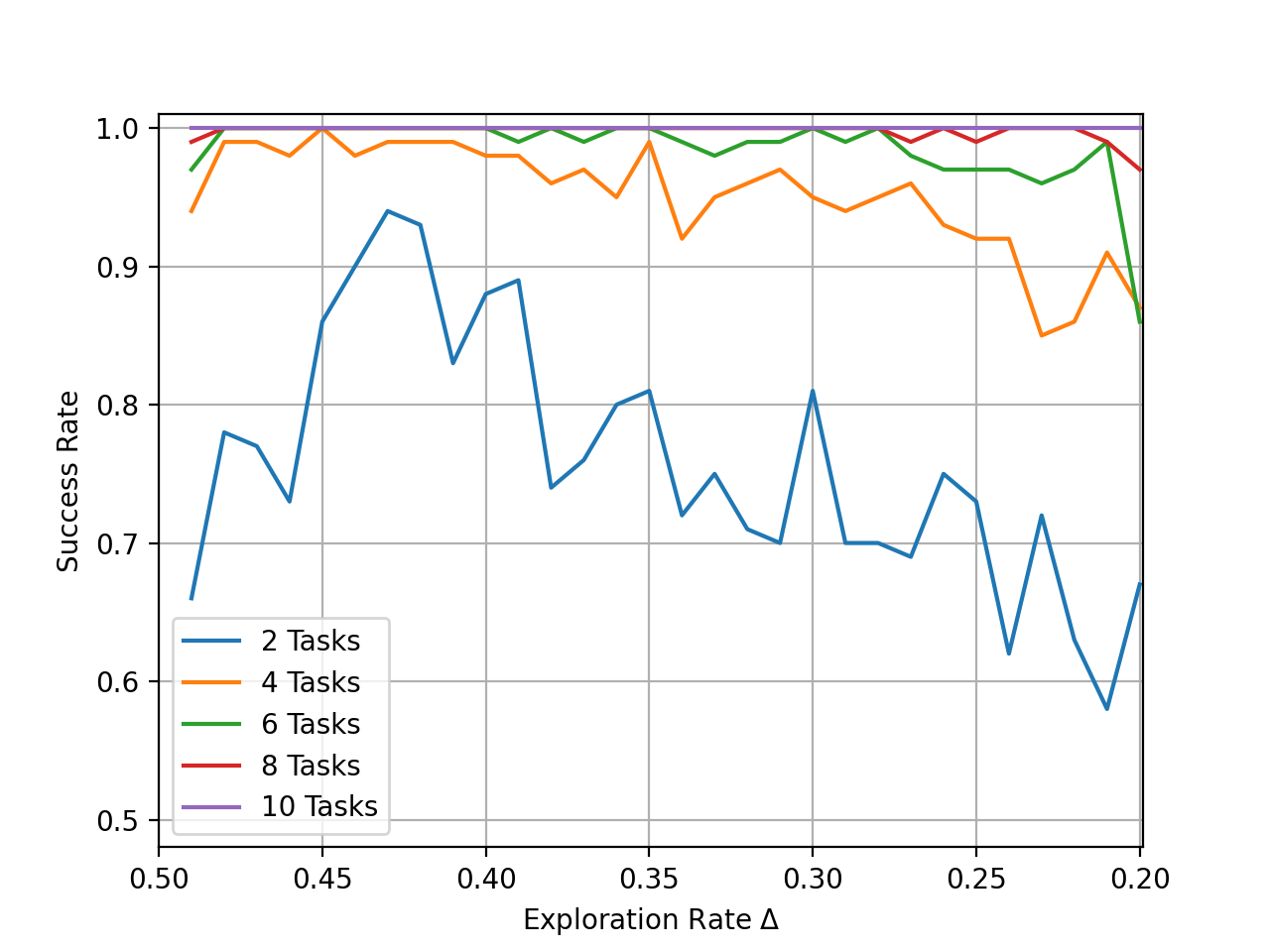}}
\begin{center}
\small \emph{(a) The success rates under different task densities}
\end{center}
\end{minipage}
\hfill
\begin{minipage}[t]{.5\linewidth}
\centerline{\includegraphics[width=\linewidth]{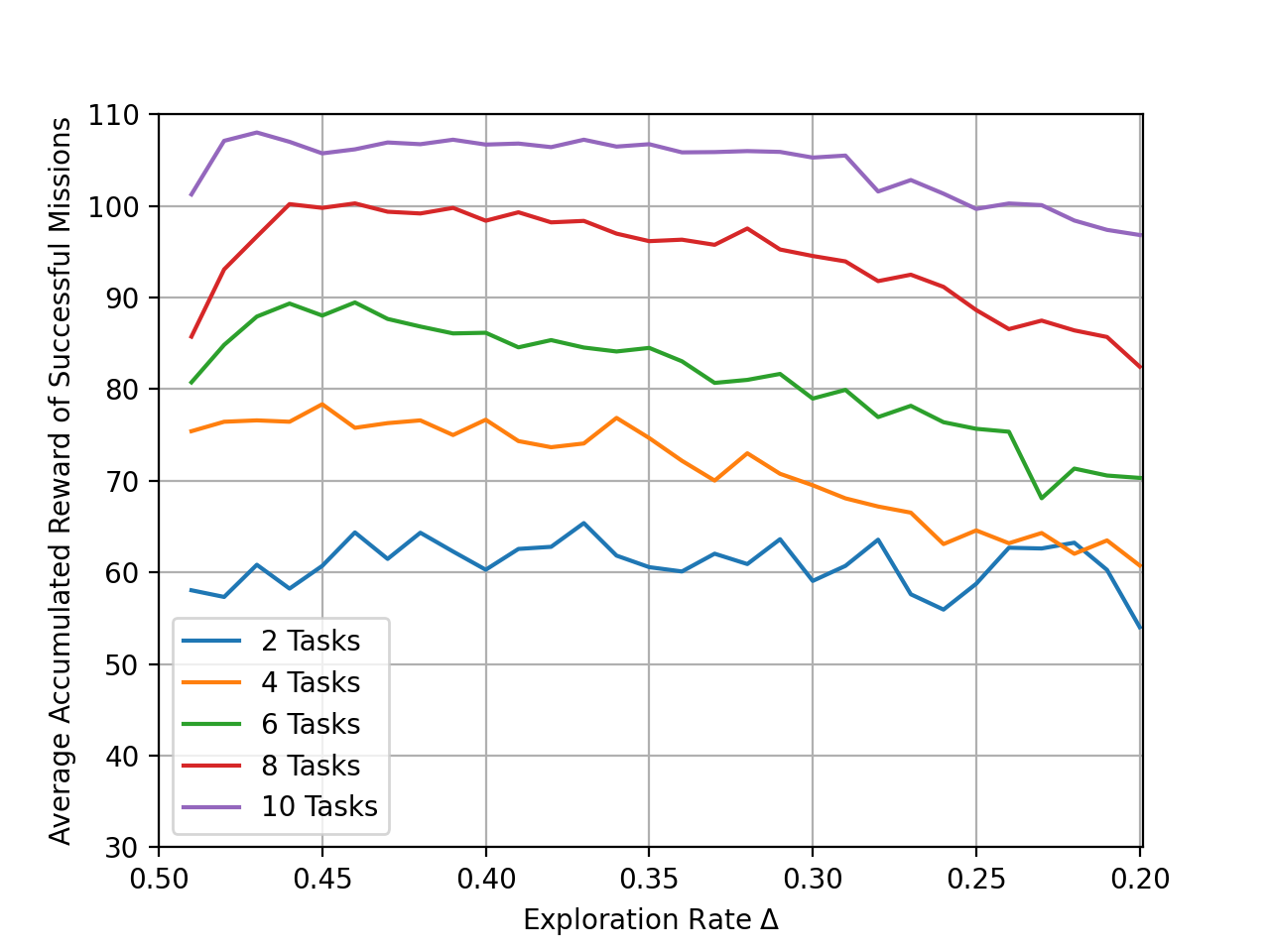}}
\begin{center}
\small \emph{(b) The average accumulated rewards of successful missions under different task densities}
\end{center}
\end{minipage}
\caption{Results for the task density experiments}
\label{fig:task density exp}
\end{figure*}

\subsection{Threshold Experiments}
In this set of experiments, we set the initial exploration rate $\Delta$ to 0.5 and gradually decrement it by $3 \times 10^{-6}$ over the experiments till it reaches 0.15. We compare the success rate at different exploration rates while varying $\psi$ following this series \{1, 2, 5, 10\}. We set the number of tasks $K = 4$, and the task length $\mathcal{T}_i = 5, \forall i \in \mathcal{K}$. The four task locations are shown as Fig.~\ref{fig:4 tasks} and remain static in all experiments of this set.


The simulation results for this set of experiments are shown in Fig.~\ref{fig:threshold exp}. From Fig.~\ref{fig:threshold exp} (a), we can observe that after learning from the first few episodes, the success rates keep comparatively steady before decreasing with the exploration rate decrement. Although the success rates of different $\psi$ values are close, the performance is relatively higher when $\psi = 5$. When $\Delta$ reaches 0.2, the success rate is still above 0.8 and the most relative to 0.9 when $\psi = 5$. The differences between the average accumulated rewards of successful missions are more pronounced, as shown in Fig.~\ref{fig:threshold exp} (b). When $\psi = 5$, the average accumulated rewards of successful missions is higher than other $\psi$ values after the initial learning phase. Therefore, we set $\psi = 5$ and let the exploration rate $\Delta$ start with 0.5 and gradually decrease to 0.2 in the other two sets of experiments.

\subsection{Task Length and Location Experiments}

We conducted three experiments in this set. The mission of each experiment is composed of four tasks, $K = 4$.  The four tasks in the first experiment locate at the trajectory points indicated by Fig.~\ref{fig:4 tasks}, and each task has a random length between one and five time steps per episode, $1 \leq \mathcal{T}_i \leq 5, \forall i \in \mathcal{K}$. The location of each task in the second experiment is set to a random potential task location indicated in Fig.~\ref{fig:4 tasks}, but every task has the same length per episode, five time steps, $\mathcal{T}_i = 5, \forall i \in \mathcal{K}$. The third experiment inspects the performance when both the location and length of each task are random per episode. The length of each task in the third experiment is a random value between one and five, $1 \leq \mathcal{T}_i \leq 5, \forall i \in \mathcal{K}$.


Fig.~\ref{fig:task len and loc exp} shows the results of this set of experiments. From Fig.~\ref{fig:task len and loc exp} (a), the success rate differences among the three experiments are subtle. Beyond the initial learning phase at the beginning of experiments, the success rates of the three experiments stay close to each other and gradually decrement to 0.8 when the exploration rate reduces to 0.2. We can learn that the task length and location don't impact the success rate significantly, even when each episode has a different set of tasks to be finished. Fig.~\ref{fig:task len and loc exp} (b) indicates that the average accumulated rewards of successful missions are higher when the task locations are random. When exploration rate $\Delta < 0.3$, the fixed task length over performs the random task lengths. We think the reason is two-fold: first, the random task location can let the proposed collaborative execution model learn to find a task more quickly, and second, the fixed task length is easier for the model to learn to collaboratively accomplish the mission than the random task length.


\subsection{Task Density Experiments}

We vary the number of tasks from two to ten in this study, and each task has a random length between one and five time steps, $1 \leq \mathcal{T}_i \leq 5, \forall i \in \mathcal{K}$, located at a random potential task location indicated in Fig.~\ref{fig:4 tasks} per episode. We define task density as the ratio between the number of tasks and the grid size. The size of the grid remains the same in this study. Hence, the larger the number of tasks, the higher the task density.

As Fig.~\ref{fig:task density exp} illustrates, the task density affects the performance of the proposed collaborative execution model. In Fig.~\ref{fig:task density exp} (a), when the task density is very sparse, e.g., two tasks in the experimental area, the success rate is much lower than higher task densities, such as four or more tasks in the same area. When the task density is high, e.g., eight or more tasks in the experimental area, the success rate is nearly 100\% regardless of the exploration rates. We believe that more tasks in the network area enable the proposed model to gain more experiences of locating tasks and learn more quickly in the early phase, allowing drones to select actions based on more accurate predictions built on those experiences. As shown in Fig.~\ref{fig:task density exp} (b), the average accumulated rewards of successful missions line up with the success rates. The average accumulated rewards of successful missions are more significant when the task density is higher, as our collaborative execution model learns better in the high task density scenarios.

\section{Conclusion and Future Work}
\label{conclusion}

We propose an energy-aware collaborative execution MARL model in this study to cope with the challenge of limited drone battery capacity in mission-oriented drone networks. We let each drone have its DQN and train it separately while sharing the same reward function across all drones to encourage them to cooperate more effectively. The battery levels of drones drive the reward function to stimulate them to learn to complete the mission collaboratively while being aware of their battery levels. 

We conducted three sets of experiments in this study to investigate the impact of some hyperparameters of the proposed model, task length and location, and task density. Based on our simulation study, our proposed model can successfully accomplish the mission assigned to the drone network at least 80\% of the time regardless of the task length and location the mission contains when the exploration rate is not less than 0.2 and the number of tasks in the experimental area is not less than four. The proposed model can achieve a 100\% success rate regardless of the exploration rates in scenarios where the task density is not too sparse.

Our next step is to evaluate our proposed model in larger grids and study the impact on its performance from the area covered by the mission. We also plan to extend the 2D space used in this study into 3D and inspect the proposed collaborative execution model in the new 3D space. We can then examine the performance of the proposed model under more extensive action and state space. Moreover, investigating and formulating the relationship between the number of drones, battery capacity, task density, and task length is also on our to-do list. Last but not least, we are working on a variation of the proposed model, in which we let each drone use an individual reward that encourages the drone to finish tasks quickly and efficiently. Our study related to this variation is three-fold: first, whether the individual reward function will affect the performance of the mission-oriented drone networks; second, whether the individual reward function will result in competition among drones; and third, whether a competitive mission-oriented drone network will over-perform a collaborative mission-oriented drone network.

\bibliographystyle{IEEEtran}
\bibliography{IEEEexample}

\end{document}